\documentclass[aip, preprint]{revtex4-1}

\usepackage{amsmath}
\usepackage{graphicx}
\usepackage{textcomp}
\usepackage[version=4]{mhchem}
\usepackage{gensymb}
\usepackage{hyperref, color}
\usepackage{multirow}
\usepackage{booktabs}
\definecolor{linkColor}{rgb}{0.8,0,0}
\definecolor{darkred}{rgb}{0.8,0,0}
\hypersetup{pdfborder={0 0 0},colorlinks=false}

\usepackage{fancyhdr}
\begin{document}

\title{Revealing the Evolution of Order in Materials Microstructures Using Multi-Modal Computer Vision}

\author{Arman Ter-Petrosyan}
\affiliation{Department of Materials Science and Engineering, University of California, Irvine, Irvine, California 92697}
\affiliation{National Security Directorate, Pacific Northwest National Laboratory, Richland, Washington 99352}

\author{Michael Holden}
\affiliation{National Security Directorate, Pacific Northwest National Laboratory, Richland, Washington 99352}

\author{Jenna A. Bilbrey}
\affiliation{National Security Directorate, Pacific Northwest National Laboratory, Richland, Washington 99352}

\author{Sarah Akers}
\affiliation{National Security Directorate, Pacific Northwest National Laboratory, Richland, Washington 99352}

\author{Christina Doty}
\affiliation{National Security Directorate, Pacific Northwest National Laboratory, Richland, Washington 99352}

\author{Kayla H. Yano}
\affiliation{Energy and Environment Directorate, Pacific Northwest National Laboratory, Richland, Washington 99352}

\author{Le Wang}
\affiliation{Physical and Computational Sciences Directorate, Pacific Northwest National Laboratory, Richland, Washington 99352}

\author{Rajendra Paudel}
\affiliation{Department of Physics, Auburn University, Auburn, Alabama 36849}

\author{Eric Lang}
\affiliation{Department of Nuclear Engineering, University of New Mexico, Albuquerque, New Mexico 87131}

\author{Khalid Hattar}
\affiliation{Department of Nuclear Engineering, University of Tennessee, Knoxville, Tennessee 37996 }

\author{Ryan B. Comes}
\affiliation{Department of Physics, Auburn University, Auburn, Alabama 36849}
\affiliation{Department of Materials Science and Engineering,  University of Delaware, Newark, Delaware 19716}

\author{Yingge Du}
\affiliation{Physical and Computational Sciences Directorate, Pacific Northwest National Laboratory, Richland, Washington 99352}

\author{Bethany E. Matthews}
\affiliation{Energy and Environment Directorate, Pacific Northwest National Laboratory, Richland, Washington 99352}

\author{Steven R. Spurgeon}
\email{steven.spurgeon@nrel.gov}
\affiliation{National Renewable Energy Laboratory, Golden, Colorado 80401}
\affiliation{Renewable and Sustainable Energy Institute, University of Colorado Boulder, Boulder, CO, 80309}
\affiliation{National Security Directorate, Pacific Northwest National Laboratory, Richland, Washington 99352}

\date{\today}


\begin{abstract}

The development of high-performance materials for microelectronics, energy storage, and extreme environments depends on our ability to describe and direct property-defining microstructural order. Our present understanding is typically derived from laborious manual analysis of imaging and spectroscopy data, which is difficult to scale, challenging to reproduce, and lacks the ability to reveal latent associations needed for mechanistic models. Here, we demonstrate a multi-modal machine learning (ML) approach to describe order from electron microscopy analysis of the complex oxide La$_{1-x}$Sr$_x$FeO$_3$. We construct a hybrid pipeline based on fully and semi-supervised classification, allowing us to evaluate both the characteristics of each data modality and the value each modality adds to the ensemble. We observe distinct differences in the performance of uni- and multi-modal models, from which we draw general lessons in describing crystal order using computer vision.

\end{abstract}

\maketitle

\section{Introduction}

Technological advancements in energy storage and quantum computing depend on our ability to design and deploy high-performance materials. A basic tenet of materials science is that function follows form: that is, a material's microstructure determines emergent properties and functionality for a given application. How we describe and design such microstructures is therefore critical to developing better materials; while conceptually simple, this task is challenging in practice. Most materials are formed or deployed in coupled extremes of temperature, pressure, radiation, and other stimuli, leading to non-linear modifications of microstructures that are difficult to predict and control.\cite{eswarappa2023materials, zhang2020ion, pernigoni2023advantages, schmidt2021effects} Our mastery of materials thus begins and ends with our ability to describe and direct microstructures.

Microstructures elude a simple definition, but in practice they usually reflect specific associations in local structure, chemistry, and composition.\cite{norouzi2024} These associations may encompass distributions of crystalline or amorphous phases, interfaces, and defects such as voids, impurities, dislocations, and stacking faults. How we measure these features depends on the material, but one of the highest resolution and most information-rich approaches is based on scanning transmission electron microscopy (STEM).\cite{Ramasse2017, Petford-Long2008} In the STEM, a convergent electron probe is scanned over a material's surface, generating a variety of modalities formed from imaging signals, spectroscopic signals, and diffraction signals that collectively inform the microstructure. The interpretation of these signals is complex and may depend on sample preparation, imaging parameters, and beam-specimen interactions.\cite{Chen2016, Spurgeon2016b} In simple mathematical terms, our goal is to reconstruct an object function from its imaging function. We do this through physics-based models specific to each signal, which reveal different characteristics of the microstructure under study. Despite recent work in data fusion,\cite{schwartz2022} there is still no widely accepted, objective metric for which STEM modalities matter and how they should be interpreted to collectively define a microstructure.

While microstructures are important across all classes of materials, here we focus specifically on complex oxides. These materials exhibit strong lattice-spin coupling that underpins useful functionality and is highly sensitive to even low concentrations of vacancies, impurities, and extended defects.\cite{gunkel2020oxygen, tuller2011point} Oxide thin-film transistors (TFTs) have emerged as key components of modern electronics.\cite{shi2021wide, geng2023thin, yan2024thin, WANG2024100396} Heterojunction oxide TFTs, in particular, demonstrate improved electrical performance as well as high operational stability.\cite{faber2017heterojunction, li2020recent, lee2023heterojunction} TFTs and other oxide-based components are commonly deployed in extreme environments such as nuclear reactors and spaceflight, where they are exposed to charged particles and photons that can introduce deleterious lattice defects,\cite{sickafus2000radiation} impacting performance and leading to early device failure. It is therefore important to quantify both intrinsic microstructures and their behavior under irradiation.

Beyond the bulk, oxide interfaces have attracted particular attention because these regions exhibit useful, distinct structural and chemical states.\cite{Stemmer2014, Zubko2011} External stimuli can uniquely couple to interfaces, leading to modified cation intermixing\cite{Yano.2022} and corrosion sensitivity.\cite{kaspar2023role, Uberuaga2021} Radiation-induced damage in particular strongly depends on interface components and their configuration.\cite{Beyerlein2015, Beyerlein2013} For instance, we have previously observed a retention of interface crystallinity upon irradiation of epitaxial \ce{LaMnO3}/\ce{SrTiO3} thin films,\cite{matthews2021percolation} and shown that asymmetries in defect formation energies strongly influence how interfaces degrade.\cite{Sassi2019, Spurgeon2020} While these and other studies\cite{Choudhury2015, Aguiar2014, Aguiar2014a, Dholabhai2014, Bi2014} have yielded valuable insights into the behavior of oxide interfaces, they are highly specialized and limited in their statistical and descriptive power. Additionally, repeatability may be low as most analysis steps are informed by researchers and therefore subject to their experience and biases.\cite{Bruno2023} To build more accurate models of interface behavior, we require more objective descriptors of microstructural evolution. Computer vision models may offer a solution to this problem by allowing for more reproducible, scalable, and informed descriptors relative to human-in-the-loop studies.

In recent years, automated methods have been developed to collect, triage, and quantify microstructural data in the microscope.\cite{kalinin2023machine, olszta2022automated} Such methods have been used to track adatom movement in 2D materials,\cite{Roccapriore2022} quantify nanoparticle morphologies,\cite{sytwu2022} identify atomic defects and dopants in monolayers,\cite{li2023single} and detect oxygen deficient regions in thin films.\cite{akers2021rapid} These methods can reproducibly and scalably analyze large volumes of data, potentially informing more accurate microstructural descriptors. While STEM is incredibly data-rich because of its various imaging, diffraction, and spectroscopy operating modes, in practice researchers only utilize a small subset of the data available to them because of their inability to deeply process hundreds or thousands of images. Furthermore, most models are painstakingly trained on single data modalities and specialized for particular materials systems, leading us to overlook important cross-modal correlations. For example, there are many approaches to image segmentation for STEM,\cite{li2021defectnet, stuckner2022microstructure, ziatdinov2022b, sadre2021deep, groschner2021machine, lee2022stem, burns2024deep, Shi2022},  but such models may not effectively discern interfaces in both structure and composition. The application of semi-supervised multi-modal models could provide insight into how different modalities influence segmentation results and, more importantly, allow us to assess whether more modalities provide a more informed microstructural descriptor.

The integrated analysis of multi-modal data streams in electron microscopy is still in its infancy.\cite{schwartz2022} In contrast, multi-modal AI/ML has found considerable use in applications of autonomous driving and remote sensing, often using a combination of cameras, Light Detection and Ranging (LiDAR), and radar.\cite{cui2022driving, rizzoli2022multimodal, jiangyun2024rs, jiang2023intelligent} Progress has been made towards integrating multiple microscopy modalities in AI/ML analyses in the life sciences,\cite{walter2020correlated, fernandez2021pyjamas, cameron2021leveraging, IBTEHAZ202074} while comparatively less work has been done in materials science \cite{muroga2023advmat}. For examples, Kimoto et al. performed a comprehensive bimodal analysis of material nanostructures using 4D-STEM image and diffraction data \cite{Kimoto2024}. In these approaches, registered images from different modalities are typically fused into multi-dimensional arrays that are passed into computer vision models. When spectral information is involved, each bin in the spectrum is treated as its own dimension to create a hyperspectral image that is then passed into the computer vision model. 

It has been recently shown that the unsupervised segmentation of digital photographs can be achieved by combining image features or learned embeddings and graph theory. Jiao et al.~applied graph clustering to high-dimensional embeddings obtained from an autoencoder which was fed low-level visual features of color, edge, gradient, and image saliency.\cite{jiao2020segmentation} Wang et al.~used graph-cuts on a graph formed from similarities between self-supervised transformer features to perform object localization.\cite{wang2022self} Melas-Kyriazi et al.~performed spectral clustering on a graph Laplacian matrix formed from a combination of color information and unsupervised deep features.\cite{melas2022deep} Aflalo et al.~incorporated node features derived from pre-trained neural networks into the graph and performed clustering using a graph neural network trained on a loss function based on classical graph clustering algorithms.\cite{aflalo2022deepcut} In addition, graph-based analytics readily support the incorporation of multiple modalities during creation of the graph representation.\cite{ektefaie2023multimodal, iyer2020graph, he2023multimodal} While graph-based image segmentation has existed prior to the widespread use of deep-learning-based segmentation,\cite{felzenszwalb2004efficient, camilus2012review} these approaches have not yet been widely applied to electron microscopy data.

In this work, we examine the multi-modal segmentation of the complex oxide interface LaFeO$_3$ (LFO)/SrTiO$_3$ (STO). We choose the LFO/STO system for its technological relevance as a potential solid oxide fuel cell (SOFC) for extreme environments.\cite{Taylor.2023} In our approach, we apply various levels of \textit{a priori} knowledge: 1) a graph-based approach that requires minimal advanced knowledge of the sample, in the form of an upper bound for similarities between identified regions, 2) an unsupervised clustering-based approach that requires one to know the expected number of discrete regions, and 3) a semi-supervised few-shot classification approach that requires an expected number of regions and a limited subset of user-selected examples of each region. In addition, we examine the effect of incorporating multiple datastreams of high-angle annular dark field (HAADF) images and energy dispersive X-ray spectroscopy (EDS) spectra on the segmentation performance. Finally, we assess how each modality effects segmentation by examining the elemental composition and crystallinity of the identified clusters. We observe distinct differences in the performance of uni- and multi-modal models, which allow us to derive latent correlations informing disordering of the LFO/STO system. More broadly, we identify important considerations and limitations of the use of multi-modal computer vision for microscopy.


\section{Results and Discussion}
LFO thin films were epitaxially grown on single-crystal STO substrates, as described in Section \ref{sec:Methods}. One set of films contained structural defects, manifested as columnar regions where local compositional homogeneity and crystallinity varied notably from the nominal stoichiometry and lattice, while the other set did not. We term these samples ``defective'' and ``pristine'', respectively. Samples from both sets were irradiated to 0.1 displacements per atom (dpa) to induce crystalline and chemical disorder. The film surface was capped with a protective layer of either Cr or Pt, and cross-sectional STEM samples were prepared. HAADF images and EDS spectra were collected, shown in Figures \ref{fig:LFOpristine} and \ref{fig:LFOdefect}. In the HAADF images, perovskite lattices are visible in the LFO and STO layers. Because the caps are polycrystalline, no clear lattice structure is present. In the EDS spectra, intensities of the Cr or Pt, Fe, and Ti peaks delimit the Cr or Pt cap, LFO, and STO layers, respectively.

To analyze relevant microstructural features, we sub-divide an image into small uniform units (or ``chips'') that can be directly compared to one another. Chipping allows one to define a scale at which meaningful structural motifs are present and evolve with irradiation.\cite{Dan2022, akers2021rapid} This step may require \textit{a priori} knowledge of expected features or processing, and further work would be needed to automate selection of appropriate motif size. In this case, we are interested in deviations from the normal periodic structure of the two perovskites, so chips should be sized accordingly to include at least one unit, on the order of 0.5--1 nm. Chipping was done across both HAADF images and EDS spectra to support multi-modal analysis.

\subsection{Comparison of Classification Approaches}
\begin{figure}[htbp]
    \centering
    \includegraphics[width=0.9\linewidth]{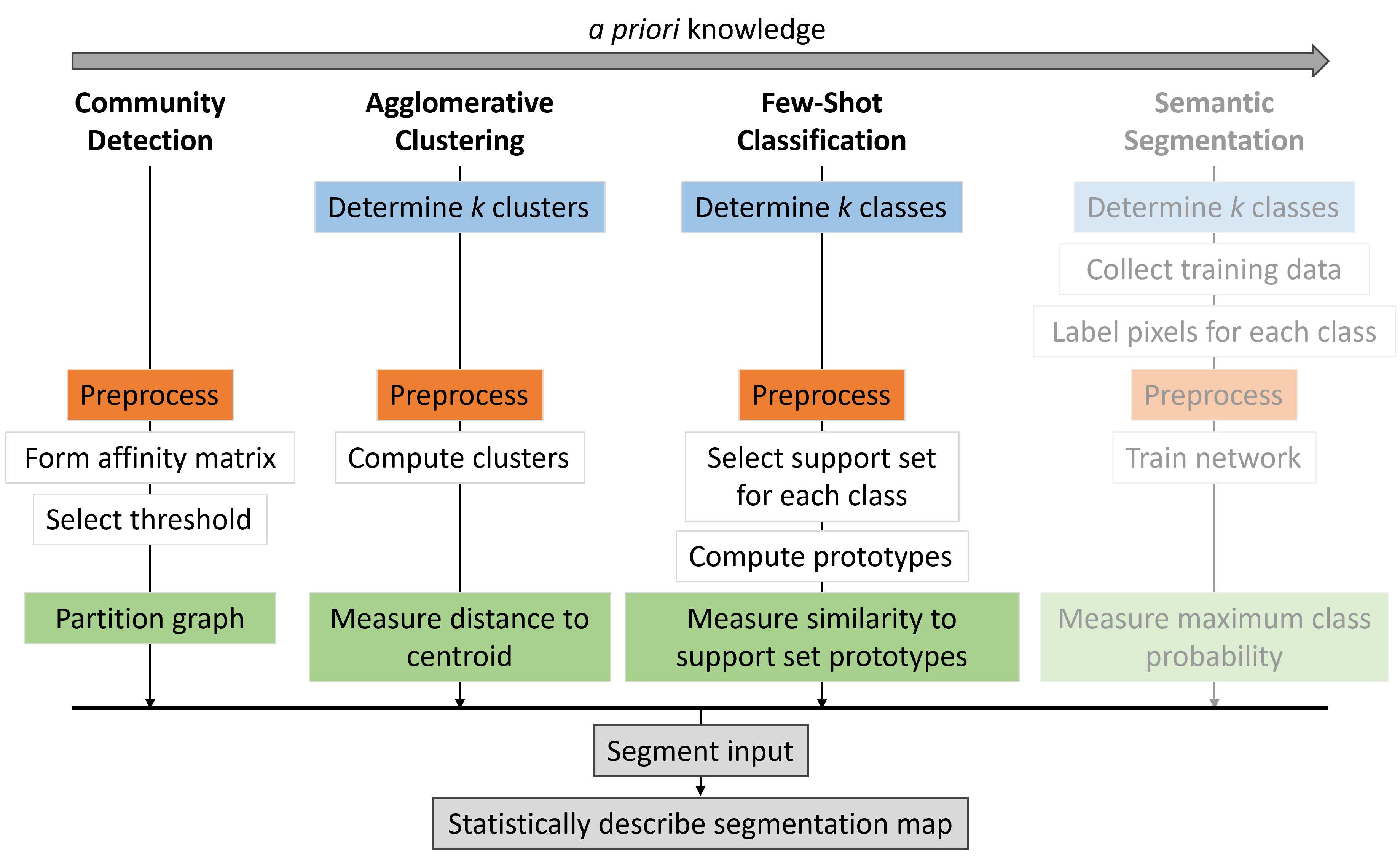}
    \vspace{-1em}
    \caption{\textbf{Methods for the segmentation of structural regions within STEM data.} The required \textit{a priori} knowledge increases by method from left to right and therefore the perceived effort, under differing trade offs, increases from left to right. Similar steps across methods are color coded to emphasize overlaps in each approach. The semantic segmentation approach is not examined in this work and only shown for comparison.}
    \label{fig:ml_workflows}
\end{figure}

We performed single- and multi-modal classification of HAADF images and EDS spectra of each sample using three different techniques: 1) community detection, 2) agglomerative clustering, and 3) few-shot classification. These three methods were chosen due to the differing amounts of \textit{a priori} knowledge required, as outlined in Figure \ref{fig:ml_workflows}. Fully supervised classification requires the most \textit{a priori} knowledge. A labeled dataset must be collected to train the classifier, which predetermines the number and identity of potential classes. Few-shot classification does not require a large labeled training set---only a small support set that defines the number and identity of potential classes is necessary. Agglomerative clustering does not require a support set, but does require the number of expected classes to be defined. Finally, community detection does not require an expected number of classes but does require an upper similarity limit to be set. This upper limit can be either a given value or a statistical measure of the graph, such as the mean, median, or mode of all chip-to-chip similarities.\cite{ter2023unsupervised} All techniques require knowledge of the expected size of microstructural features, as this information determines the chip size.

The degree of \textit{a priori} knowledge has a subtle influence on the outcome of each approach. Community detection, requiring no \textit{a priori} knowledge, tends to produce outliers or one-off clusters as an artifact of strictly data driven boundaries, but, as it requires no user input, it tends to produce more repeatable results. Agglomerative clustering incurs a slight increase in \textit{a priori} requirements, i.e. defining the number of clusters, and resolves outliers into fewer groupings. However, this reduction of groups may potentially dilute the physically meaningful qualities of the cluster. For few-shot classification, the support set selection relies heavily on \textit{a priori} domain knowledge of the material system. Determining which and how many chips to designate as support examples will directly influence how representative the prototypes are and how physically meaningful the model outcomes will be.\cite{snell2017prototypical} Lastly, the fully supervised system demands detailed annotations for large volumes of training data, rooted in labor intensive labeling and complete \textit{a priori} knowledge. This approach takes advantage of fully described systems to form knowledge-driven decision boundaries and holds the potential for high fidelity outcomes. However, these labor intensive and high fidelity systems are often brittle, failing to generalize outside of training distribution data. Because we do not have a large, previously collected dataset describing irradiation-induced disorder in epitaxially grown films, we do not explore fully supervised methods in this work.

\begin{figure}[htbp]
    \centering
    \vspace{-4pt}
    \includegraphics[width=0.45\textwidth]{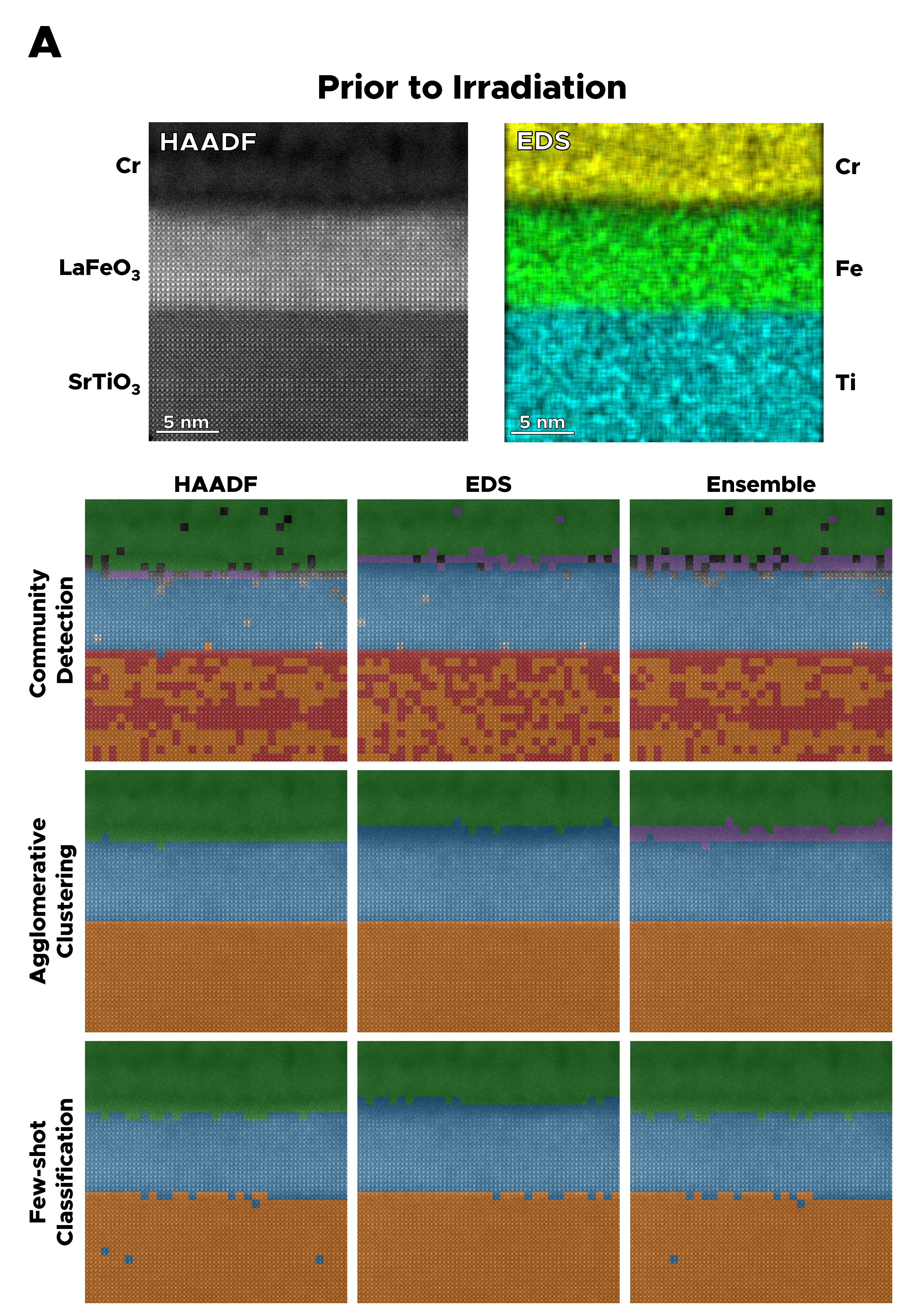}
    \includegraphics[width=0.45\textwidth]{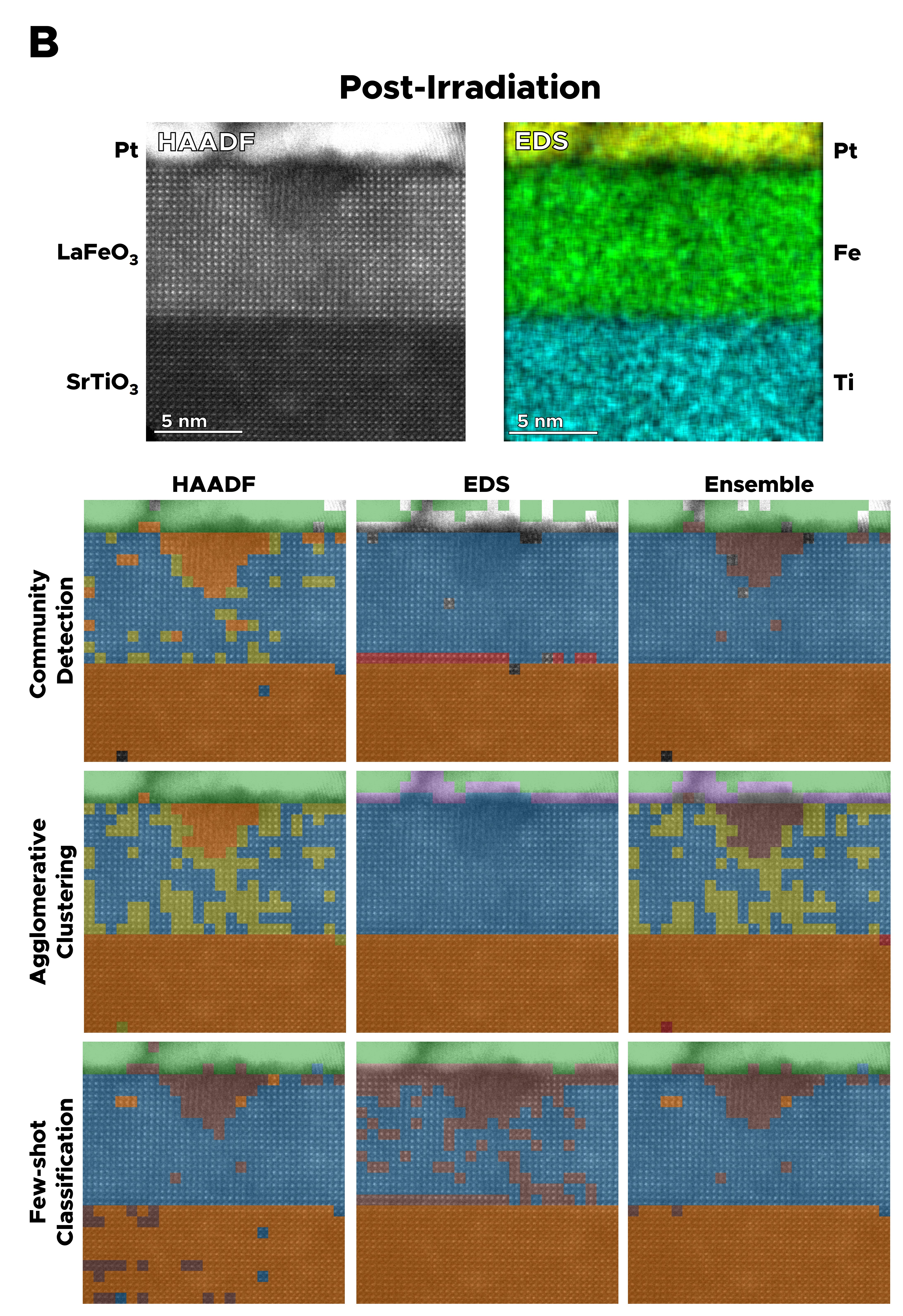}
    \vspace{-14pt}
    \caption{\textbf{(A--B) Clustering of pristine LFO/STO sample before and after irradiation to 0.1 dpa, respectively.} Clustering based on single modalities can distinguish the initial film layers, but not the triangular amorphous region induced by irradiation. For the latter case, a multi-modal approach exhibits improved segmentation.}
    \label{fig:LFOpristine}
\end{figure}

Multiple data modalities can be incorporated in segmentation analyses in a variety of ways. Here, we use ensembling to accomplish multi-modal clustering. During ensembling, the classification results of each modality are joined, and the resulting ordered vector is used as the class label. This method is reminiscent of a majority voting ensemble. We chose ``vector voting'' instead of a traditional ensembling method to accommodate ensembling of two modalities. Hard voting ensembles use the mode for classification, which is not applicable when the two modalities disagree. Soft voting ensembles average the class probabilities and assigns the class based on the highest weighted probability. While few-shot classification does produce probabilities for each class, the applied clustering techniques do not.
PTe few-shot classification and agglomerative clustering techniques  demonstrated consistent performancepcross eallmodalityies prior to irradiation, ,sas hown in Figure \ref{fig:LFOpristine}.A for the pristine sample. We find intelligible segmentation of the sample image that aligns with the STO, LFO, and Cr cap layers observable in the HAADF image and EDS maps. Community detection and ensemble agglomerative clustering identify an additional cluster (shown in purple) that corresponds with the interfacial region between the LFO layer and Cr cap. The community detection approach segments the STO layer into two regions (shown in red and orange) for both individual modalities and the ensemble. The regions do not appear to correspond to any identifiable feature in the HAADF image and do not appear to have different elemental compositions, which most likely represents an artifact based on the lack of \textit{a priori} information.

\begin{figure}[h]
    \centering
    \vspace{-4pt}
    \includegraphics[width=0.45\textwidth]{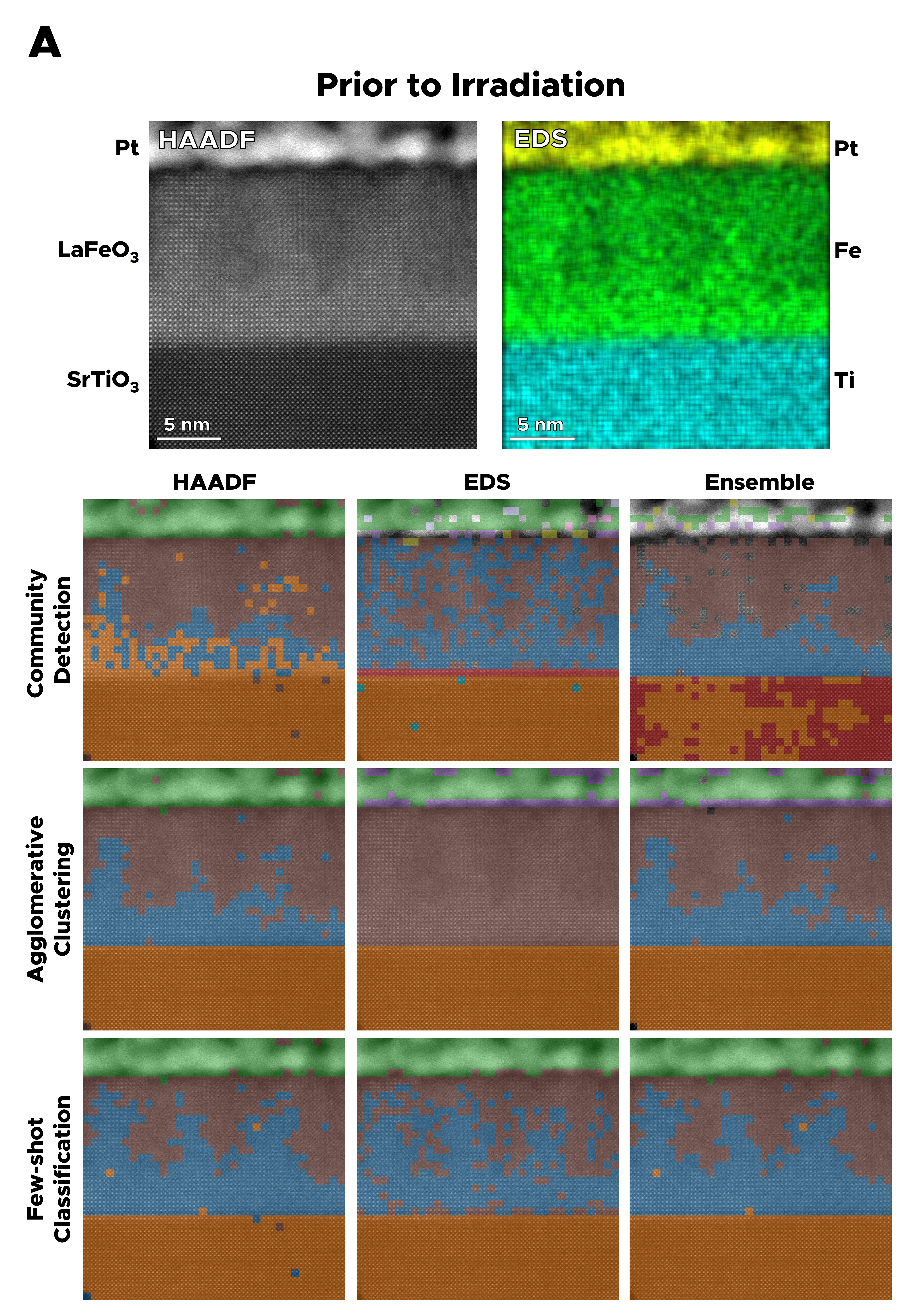}
    \includegraphics[width=0.45\textwidth]{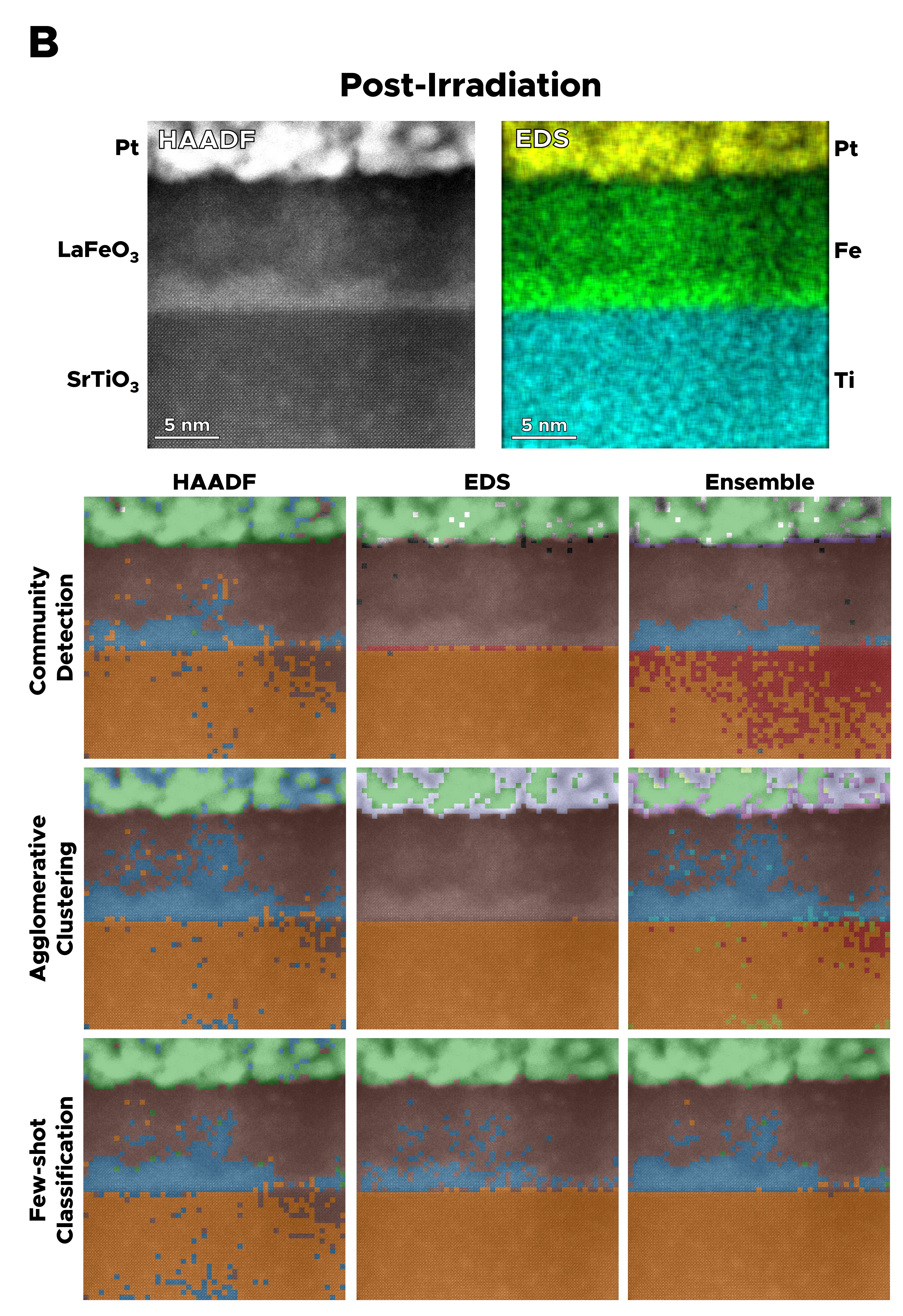}
    \vspace{-14pt}
    \caption{\textbf{(A--B) Clustering of defective LFO/STO sample before and after irradiation to 0.1 dpa, respectively.} Clustering based on a single modality fails to adequately distinguish the structural defects and their evolution, whereas the ensemble approach can more readily distinguish these defects.}
    \label{fig:LFOdefect}
\end{figure}

In the case of the defective sample prior to irradiation, shown in Figure \ref{fig:LFOdefect}.A, the segmentation is more challenging. Specifically, the community detection approach fails to distinguish the vertical striations delineating regions between defect domains in the LFO from the STO layer, thereby introducing substantial noise in the labeling of the STO layer. While the agglomerative clustering approach detects the defect domains on the basis of the HAADF image, it is not able to do so with the EDS data alone, and instead the best performance is achieved with the ensemble approach. The few-shot model, on the other hand, demonstrates the best performance distinguishing the domains within each individual modality (apart from a few outlier chips) and its ensemble performance is similar to the agglomerative clustering ensemble approach.

Post irradiation, the three techniques display significant performance differences, as shown in Figure \ref{fig:LFOpristine}.B for the pristine sample. All techniques successfully segmented the sample image into three regions roughly corresponding to the STO, LFO, and cap layers seen in the HAADF image. Community detection and agglomerative clustering failed to identify the LFO disorder when only EDS was used for clustering, likely due to the high similarity between the elemental composition of the crystalline and amorphous regions. Rather, these approaches grouped the region with the STO layer when only HAADF was used, likely due to the similar contrast in those regions of the image. Instead of surrounding the most-visible disordered region in LFO, the fourth cluster (shown in yellow) contains chips located in the LFO layer. In the agglomerative clustering case, this cluster surrounds the expected region of disorder. These chips may be associated with transition from irradiation-induced disorder to the ordered crystal. Ensembling resolves the LFO defect region for both approaches, but the subtle transition cluster only remains in the agglomerative clustering approach. Few-shot classification successfully identified and differentiated the disordered region near the top of the LFO layer across all modalities, including when only EDS was used, which indicates that subtle changes in elemental composition can be detected when the approach is given enough information.

For the post-irradiated defective sample, clusters corresponding to the defective region were produced, as shown in brown in Figure \ref{fig:LFOdefect}.B, but the size of the defective region was inconsistent between modalities and clustering techniques. For each technique, the use of only HAADF images led to significant noise, in terms of the spatial scattering of cluster assignments. When only EDS spectra are used, both community detection and agglomerative clustering failed to distinguish the LFO crystalline regions, while the few-shot technique identified the most visually prominent crystalline regions. For the ensemble approach, agglomerative clustering and few-shot classification showed reduced noise compared to the HAADF-only clusters, while recovering crystalline regions absent from the EDS-only clusters. Since the ensemble encodes both imaging and spectroscopic data, it is most effective at distinguishing the changes in signal that result from irradiation, which is expected to introduce amorphization, changes in chemical state, and migration of alloying elements.

\subsection{Extracting Physical Descriptors}

\begin{figure}[h]
    \centering
    \vspace{-4pt}
    \includegraphics[width=0.9\textwidth]{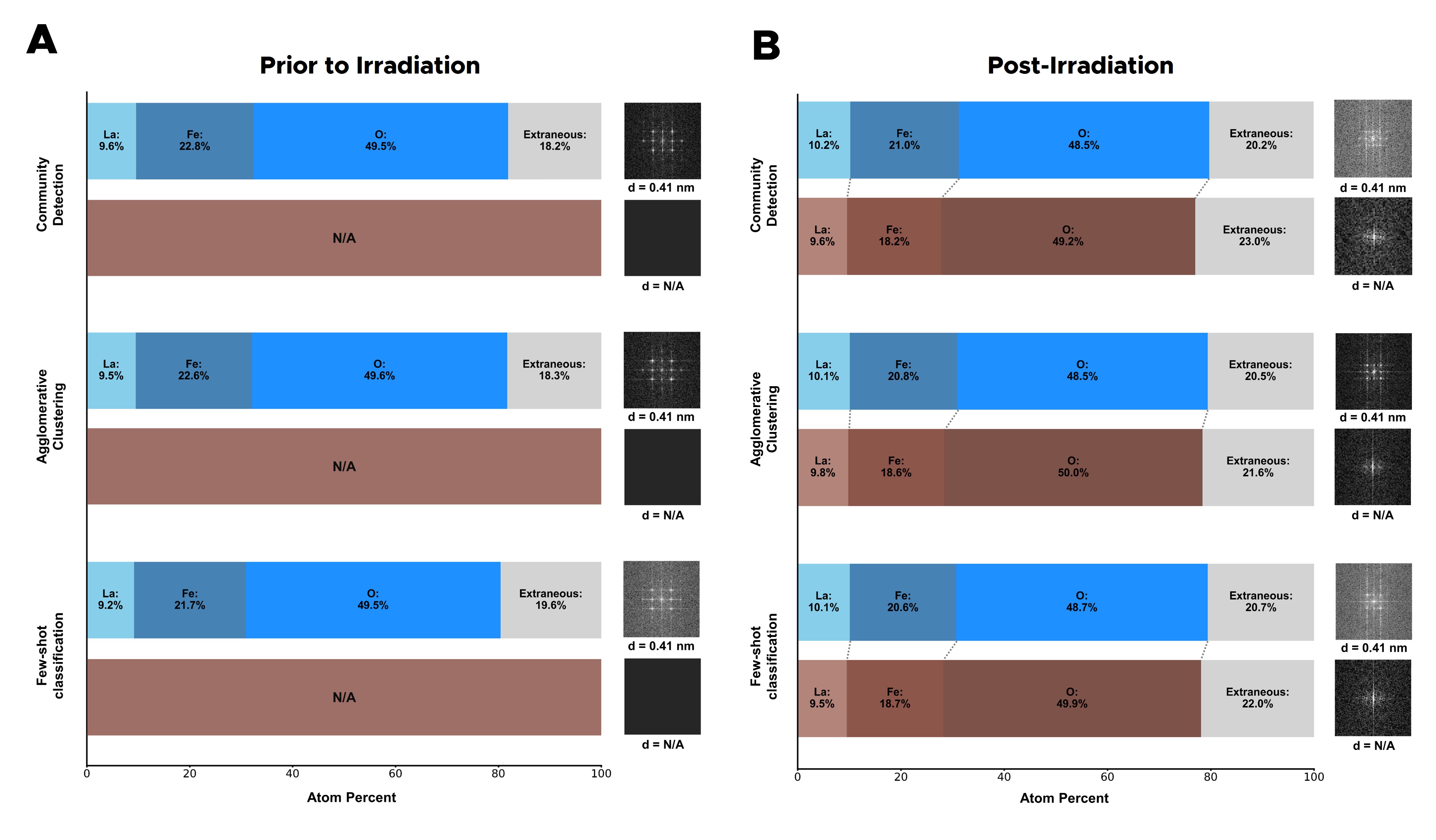}
    \vspace{-14pt}
    \caption{\textbf{(A--B) Composition and FFTs of identified regions in the pristine LFO layer before and after irradiation to 0.1 dpa, respectively.} Blue bars represent the ordered region, while brown bars represent the disorder region. Differences in the irradiated film indicate that the disordered regions contain a relatively higher local concentration of oxygen compared to the ordered regions.}
    \label{fig:LFOpristine-elems}
\end{figure}

\begin{figure}[h]
    \centering
    \vspace{-4pt}
    \includegraphics[width=0.9\textwidth]{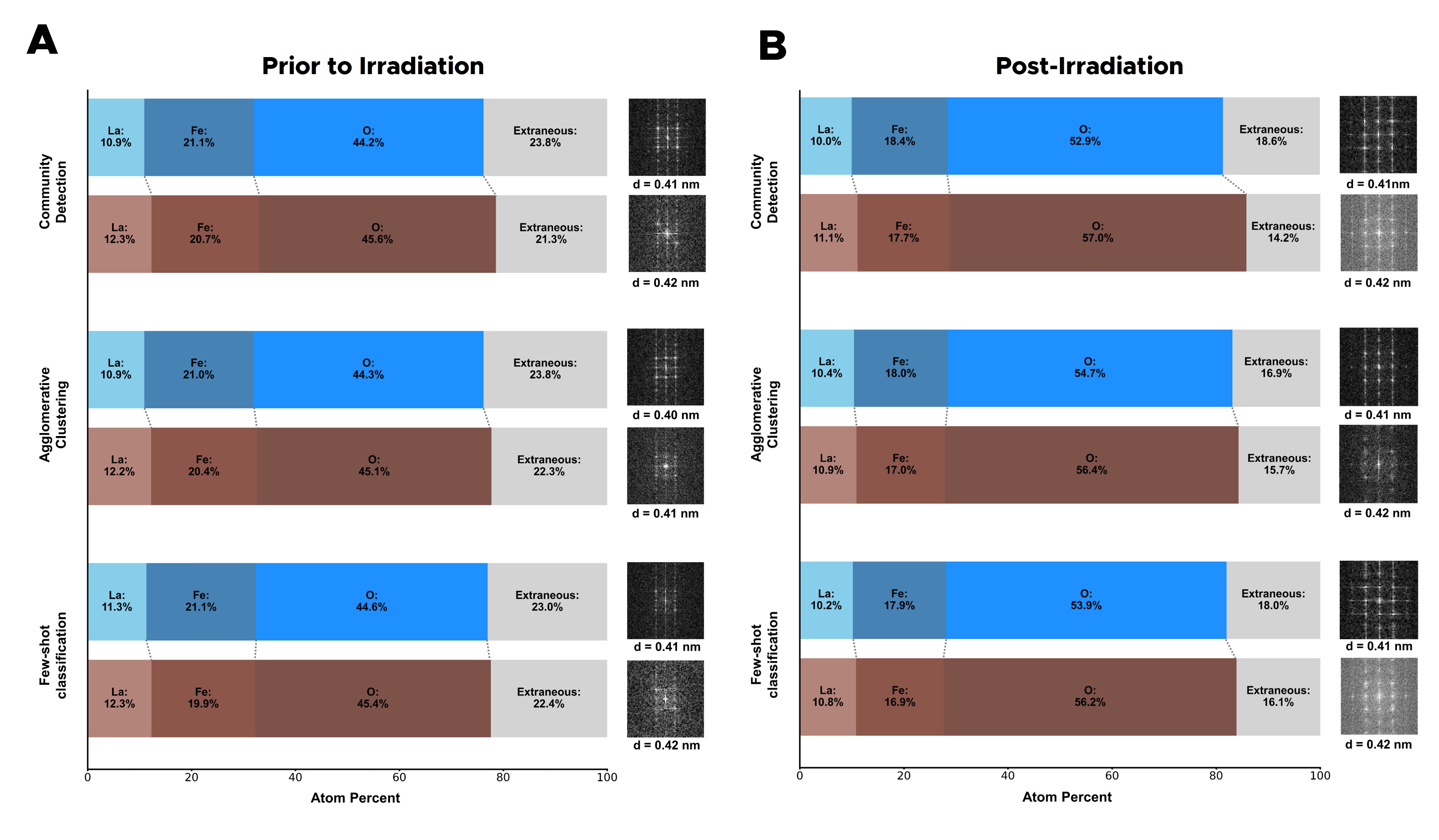}
    \vspace{-14pt}
    \caption{\textbf{(A--B) Composition and FFTs of identified regions in the defective LFO layer before and after irradiation to 0.1 dpa, respectively.} Blue bars represent the ordered region, while brown bars represent the disorder region. The intrinsic disordered regions exhibit higher local oxygen content relative to the ordered regions, which substantially increases upon irradiation.}
    \label{fig:LFOdefect-elems}
\end{figure}

We next address the topic of explainability by performing a post-clustering analysis of the observed correlations. We focus on crystallinity as measured by fast Fourier transforms (FFTs) and local elemental composition derived from EDS. These measurements allow us to extract key physical descriptors of the crystal under irradiation. Due to the variability in EDS results introduced by factors such as sample preparation, k-factor calculation, detector calibration, and beam-sample interaction,\cite{Williams2009} we focus exclusively on intra-sample comparisons to ensure more reliable and consistent analysis. Percent differences were calculated to provide a normalized metric for comparing changes in the atomic percent values of each element: La, Fe, and O. To be sure all elemental signal was accounted for when calculating atomic percent, all other signal (including background signal from the microscope column, carbon from hydrocarbons, etc.) are included under the label ``Extraneous.'' Figures \ref{fig:LFOpristine-elems} and \ref{fig:LFOdefect-elems} show the chemical composition and $d$-spacing (lattice spacing) values for the LFO order/disorder regions categorized via each clustering method in both pristine and defective samples. The pristine sample (Figure \ref{fig:LFOpristine-elems}.A) has no discernible defect domains or disordered regions and acts as a baseline to demonstrate that each clustering method does not erroneously detect disorder regions in the sample. The ordered LFO regions return a $d$-spacing value of 0.41 nm, and the elemental compositions are within a margin of error across all three clustering methods. This result lends more legitimacy to the disorder region predictions in the irradiated and defect domain samples; we can be confident that the models are not detecting random patterns in the data as defects.

For the irradiated pristine sample (Figure \ref{fig:LFOdefect-elems}.B), the ordered LFO region for each clustering method exhibits a consistent $d$-spacing value of 0.41 nm. The $d$-spacing of the disordered regions of the LFO could not be measured due to the amorphous crystal structure, which is characteristic of irradiation-induced damage in LFO. The elemental compositions of the order and disorder regions are within the margin of error across all three clustering methods. However, when comparing the chemical composition between the ordered and disordered regions, a slight increase in oxygen content was observed in the disordered region, with percent differences of 1.4\%, 3.1\%, and 2.4\% for the community detection, agglomerative clustering, and few-shot methods, respectively. Correspondingly, La content decreased by 6.1\%, 3.0\%, and 6.1\%, while Fe showed more substantial decreases, with a percent difference of 14.3\%, 11.2\%, and 9.7\% across the corresponding clustering methods.

For the defective sample prior to irradiation (Figure \ref{fig:LFOdefect-elems}.A), the ordered LFO regions exhibit a $d$-spacing value of 0.41 nm for the community detection and few-shot methods and 0.40 nm for the agglomerative clustering. The $d$-spacing values for the disordered regions are slightly larger across all three methods (0.42 nm, 0.41 nm, and 0.42 nm for the community detection, agglomerative clustering, and few-shot methods, respectively). The regions with increased $d$-spacing correlate with the defect domains which disrupt the periodic crystallographic structure within the sample. The elemental compositions are almost identical and fall within the margin of error across all three clustering methods for each: the ordered and disordered regions. Comparison of the chemical composition between the ordered and disordered regions once again shows a slight increase in oxygen content in the disorder region, with differences of 3.1\%, 1.8\%, and 1.8\% detected using the community detection, agglomerative clustering, and few-shot methods, respectively. This increase was accompanied by  decreases in La content, exhibiting a percent difference of 12.1\%, 11.3\%, and 8.5\%, and minor reductions in Fe content of 1.9\%, 2.9\%, and 5.9\%, across the three clustering methods.

The defective sample that was subjected to irradiation (Figure \ref{fig:LFOdefect-elems}.B) shows a $d$-spacing of 0.41 nm in the ordered region for the community detection, few-shot, and agglomerative clustering methods, while the disordered region has a slightly larger $d$-spacing, measuring 0.42 nm. The elemental compositions of the order and disorder regions across the three clustering methods are less consistent compared to other samples but remain within a small margin of error. This additional deviation can be attributed to variations in the predicted regions across each method, as illustrated in the ensemble prediction plots on the right side of Figure \ref{fig:LFOdefect}.B. Finally, when comparing the chemical composition between the ordered and disordered regions, the disordered region exhibits a relative increase in oxygen content and a decrease in La and Fe content. Specifically, the percent differences in oxygen are 7.5\%, 3.1\%, and 4.2\% for the community detection, agglomerative clustering, and few-shot methods, respectively. La decreases by 10.4\%, 4.7\%, and 5.7\%, while Fe decreases by 3.9\%, 5.7\%, and 5.7\% for the same methods.

Taken together, our multi-modal analyses reveal several key trends that can inform physical models. Specifically, we observe that under irradiation local O concentration increases in disordered regions similarly for both the pristine and defect-domain samples, but is slightly elevated for the irradiation of the latter. The irradiated pristine sample shows a larger decrease in local Fe content compared to La content. Conversely, the defect-domain samples show a different trend, with a larger decrease in local La content than Fe prior to irradiation, and a similar decrease in Fe and La after irradiation, which may correlate to the initial presence of defect domains and associated O inhomogeneity.

\subsection{Evaluating Model Performance}

Though quantitative evaluation of model performance absent a labeled ground truth is challenging, we can compare the three clustering approaches to evaluate their relative merits. The adjusted mutual information (AMI) score provides a metric by which to compare classification results of pairs of modalities, with values closer to 1 indicating more similar cluster membership between the two modalities. As shown in Table \ref{tab:AMI}, for the pristine sample prior to irradiation, the individual HAADF and EDS modalities produce highly similar membership for agglomerative clustering (AMI = 0.86) and few-shot classification (0.80) and moderately similar membership for community detection (0.61). Comparison of the individual modalities against the ensemble results shows that the HAADF modality contributed more to clustering via community detection and few-shot classification, while the individual modalities contributed roughly equally in agglomerative clustering. 

Post-irradiation, the AMI scores for the pristine sample between individual modalities decrease for all techniques, indicating an increased difference between the modalities. Interestingly, the individual modalities contributed similarly to ensembling during community detection, while agglomerative clustering and few-shot classification showed much higher similarity between the ensembling results and HAADF clustering results (0.90 and 0.87, respectively) than the EDS clustering results (0.79 and 0.71, respectively). In none of the cases did the EDS modality contribute significantly more than the HAADF modality.

\begin{table}[h]
\caption{Adjusted mutual information (AMI) scores between modalities for pristine and defective LFO/STO before and after irradiation.}
\begin{tabular}{ccccc}
\hline
\bf{Sample} & \bf{Modalities} & \begin{tabular}[c]{@{}c@{}}\bf{Community}\\ \bf{Detection}\end{tabular} & \begin{tabular}[c]{@{}c@{}}\bf{Agglomerative}\\ \bf{Clustering}\end{tabular} & \begin{tabular}[c]{@{}c@{}}\bf{Few-shot}\\ \bf{Classification}\end{tabular} \\ \hline
\multirow{3}{*}{\begin{tabular}[c]{@{}c@{}}Pristine\\ 0 dpa\end{tabular}} & HAADF-EDS & 0.61 & 0.86 & 0.80 \\
 & HAADF-Ensemble & 0.86 & 0.94 & 0.97 \\
 & EDS-Ensemble & 0.69 & 0.93 & 0.81 \\ \hline
\multirow{3}{*}{\begin{tabular}[c]{@{}c@{}}Pristine\\ 0.1 dpa\end{tabular}} & HAADF-EDS & 0.55 & 0.64 & 0.62 \\
 & HAADF-Ensemble & 0.73 & 0.90 & 0.87 \\
 & EDS-Ensemble & 0.74 & 0.79 & 0.71 \\ \hline
\multirow{3}{*}{\begin{tabular}[c]{@{}c@{}}Defective\\ 0 dpa\end{tabular}} & HAADF-EDS & 0.48 & 0.79 & 0.70 \\
 & HAADF-Ensemble & 0.48 & 0.96 & 0.96 \\
 & EDS-Ensemble & 0.50 & 0.84 & 0.72 \\ \hline
\multirow{3}{*}{\begin{tabular}[c]{@{}c@{}}Defective\\ 0.1 dpa\end{tabular}} & HAADF-EDS & 0.57 & 0.60 & 0.61 \\
 & HAADF-Ensemble & 0.65 & 0.86 & 0.76 \\
 & EDS-Ensemble & 0.68 & 0.80 & 0.81 \\ \hline
\end{tabular}
\label{tab:AMI}
\end{table}

Prior to irradiation, all techniques identify the defects located in the LFO layer when using HAADF data. Only the agglomerative clustering method does not identify the defects when using EDS data, though this is resolved in the ensembling results. The community detection approach using HAADF data clusters part of the LFO defect with the bottom STO layer, but again this is resolved with ensembling. As observed in the pristine film, the community detection approach ensemble identified two regions in the STO layer, though this time only during ensembling.

Similar to the results obtained prior to irradiation, all the techniques identify a new defect found in the LFO layer when using HAADF data from the post-irradiated sample. However, only few-shot classification was able to pick up these defects using only EDS data. For all three techniques, ensembling located the defects, though at varying amounts. Again, community detection separates the STO layer into two major regions, shown in orange and red in Figure \ref{fig:LFOdefect}. Close examination of the HAADF image shows a slight contrast change in the regions colored red, which is likely a result of thickness changes across the lamella and/or damage to the STO layer resulting from irradiation.

The greater utility of the HAADF images is also borne out in the AMI scores in Table \ref{tab:AMI}. Before irradiation, the HAADF-only and ensemble results are highly similar for agglomerative clustering and few-shot classification, while the community detection ensemble results are intermediate to the two single modalities. After irradiation, the contributions from HAADF and EDS become more comparable (within 0.06). These results suggest the importance of weighting modalities, depending on the segmentation task, and merits future study.

\section{Conclusions}

While the value of multi-modal data may be naively apparent, our study indicates that multi-modal data analysis is not trivial. A comparison of several clustering approaches reveals the challenges of integrating such data absent a ground truth. In general, we find that an ensemble approach incorporating community detection is the most well suited to combined HAADF imaging and EDS spectroscopy. By analyzing FFTs of the resulting clusters and quantifying EDS spectra, we are able to discriminate relative changes in local crystallinity and alloying element behavior. We observe changes in the distribution of oxygen content and loss of crystalline order, which is expected given the propensity for these materials to form oxygen vacancies.

Multi-modal descriptors allow us to more uniquely and quantitatively describe radiation-induced disorder, which will inform future kinetic models. More broadly, the use this approach taps into the wide variety of data that collectively inform physical mechanisms that would otherwise be difficult. This information will aid the execution of autonomous experiments,\cite{olszta2022automated, creange2022towards} particularly in high-speed or low-dose scenarios where additional modalities may come at great cost. However, future work is needed to better quantify model accuracy---perhaps on the basis of synthetic data---and provide guidance to emerging autonomous systems on how many and what modalities of data to acquire. Nonetheless, robust multi-modal models promise to greatly enhance our understanding of latent physical correlations in materials and chemical systems.

\section{Methods}\label{sec:Methods}

\subsection{Sample Growth and Irradiation}
Thin films of LaFeO$_3$ (LFO) were epitaxially grown approximately 10--15 nm thick on (001)-oriented SrTiO$_3$ (STO) by two methods: pulsed laser deposition (PLD) and molecular beam epitaxy (MBE). PLD films were grown to incorporate vertical defect domains. Growth parameters for the PLD samples are outlined by Stoerzinger et al.,~\cite{stoerzinger2018linking} and growth parameters for the MBE films are described by Burton et al.~\cite{burton2022thickness}

Films were irradiated to a damage level of 0.1 displacements per atom (dpa) ($3.74 \times 10^{13}$ ions/cm$^2$) at nominally room temperature at the Ion Beam Laboratory at Sandia National Laboratories, with 2.8 MeV Au$^{2+}$, at 3\degree~off normal to minimize channeling effects and a flux of approximately $3.7 \times 10^{11}$ ions/cm$^2$/sec. Dose was calculated using the Stopping Range of Ions in Matter (SRIM-2013) Monte Carlo approach to replicate experimental conditions and ensure the Au implantation peak was within the substrate.~\cite{ziegler2013srim} A mask was used to cover sections of the films during irradiation to maintain pristine portions.

\subsection{STEM/EDS Measurements}
Lamellae from each region (0 and 0.1 dpa) from both films (PLD and MBE grown) were extracted by standard focused ion beam (FIB) procedures and examined by STEM imaging and EDS. Imaging and spectroscopy were conducted on a probe-corrected 300 kV Thermofisher Themis Z with a beam convergence of 18 mrad and a collection angle range of 50 to 200 mrad (for the HAADF detector). A dwell time of 2 \textmu s was used to collect spectra in $512 \times 512$ arrays for EDS mapping, resulting in pixel sizes between 35--71 pm. 

\subsection{Data Pre-Processing}
HAADF images were chipped into small, non-overlapping, uniform units depending on the scale of the meaningful microstructural features. Table \ref{tab:chip} gives the magnification and resultant chip sizes for each sample examined in this work. EDS data were processed with the open-source Python library HyperSpy.\cite{francisco_de_la_pena_2024_12724131} The data was segmented into chips corresponding to the chipping of the HAADF image. The EDS spectra were summed to produce a single spectrum per chip. Atomic percents of Al, C, Cr, Cu, Fe, Ga, La, N, O, Pt, Sr, and Ti were calculated using the Cliff-Lorimer method.

\begin{table}[h]
\caption{Chipping information for each sample.}
\begin{tabular}{lcccccc}
\hline
 \bf{Sample}  & \bf{Dose (dpa)} & \bf{Magnification (Mx)} & \bf{Chips} & \bf{Rows} & \bf{Columns} & \bf{Chip Size} \\ \hline
Pristine & 0.0 & 3.90 & 1089 & 33 & 33 & 15 \\
Pristine & 0.1 & 5.50 & 576 & 24 & 24 & 20 \\
Defective & 0.0 & 3.90 & 1156 & 34 & 34 & 15 \\
Defective & 0.1 & 2.75 & 2601 & 51 & 51 & 10 \\
\hline
\end{tabular}
\label{tab:chip}
\end{table}

Each HAADF chip was fed into a VGG16 model pretrained on the ImageNet dataset, and the learned embedding from the final convolutional layer was extracted to produce a 512-length descriptor.\cite{simonyan2015} Chipped EDS spectra (length 4051) were used directly for clustering in some cases, while in others the vector of atomic percents (length 11) for the quantified spectra were used.

\subsection{Few-Shot Classification}
We expanded upon our previously developed few-shot approach\cite{akers2021rapid} to classify distinct regions within each sample through comparison to a small set of hand-selected prototype chips, hereafter referred to as the support set. In this approach, the Euclidean distances between the chip embeddings and the mean embedding of each prototype set are computed. A probability distribution is computed over the chip-to-prototype distances using the \textit{softmax} function. The class of the prototype with the largest probability is then assigned to the chip. Support sets for each sample were chosen by examining the HAADF image. Support sets were made to describe crystalline regions of the STO layer, crystalline regions of the LFO layer, amorphous regions of the LFO layer, and the Cr capping layer. Each set consisted of four chips.

Multi-modal ensembling was obtained by averaging the predictive scores of each modality, except for the irradiated sample with defects, where scores were combined using a 55\% EDS and 45\% image weighting scheme.

\subsection{Agglomerative Clustering}
Agglomerative clustering is a hierarchical clustering technique that uses a similarity measure to recursively merge pairs of clusters. In this work, we used the agglomerative clustering method implemented in scikit-learn.\cite{pedregosa2011scikit} The Euclidean distance between chip embeddings was used as the similarity measure, and Ward's method\cite{ward1963hierarchical} was used to determine when pairs of clusters were merged. The clustering was considered complete when the specified number of clusters is reached. 

\subsection{Graph Clustering} 
Louvain community detection, like agglomerative clustering, is a hierarchical technique that recursively merges groupings of chips. However, instead of directly using the pairwise similarities to determine when merging should occur, the modularity of each community is maximized to produce groupings with high intraconnectivity. The affinity matrix used for community detection was formed by computing the cosine similarities between all pairs of chips in a sample. The resulting affinity matrix represents a complete graph with chips as nodes and edges connecting nodes weighed by the corresponding cosine similarity. The graph connectivity was then reduced by filtering edges with weights below a certain threshold to enable the formation of distinct communities for identification by Louvain community detection. Here, we used the mode of the similarity scores as the filtering threshold. Detailed examination of the choice of filtering threshold can be found in our previous work.\cite{ter2023unsupervised} In the multi-modal scenario, the affinity matrices of each modality are averaged. After filtering the averaged graph, Louvain community detection was applied.

\subsection{Post-Segmentation Analysis}
After clustering, the composition of each cluster in the LFO region was quantified using HyperSpy in the same manner as during pre-processing. Quantification was performed over the sum of all spectra in the cluster. The interatomic spacing for each cluster was calculated by performing a FFT on the largest square region in the HAADF image to generate a reciprocal space projection and determine the d-spacing of the predicted region.

\clearpage

\section{Acknowledgements}
This work was supported by the Laboratory Directed Research and Development (LDRD) program at Pacific Northwest National Laboratory (PNNL). PNNL is a multiprogram national laboratory operated for the U.S. Department of Energy (DOE) by Battelle Memorial Institute under Contract No. DE-AC05-76RL0-1830. Some sample preparation was performed at the Environmental Molecular Sciences Laboratory (EMSL), a national scientific user facility sponsored by the Department of Energy's Office of Biological and Environmental Research and located at PNNL. Ion irradiation work was performed, in part, at the Center for Integrated Nanotechnologies, an Office of Science User Facility operated for the U.S. Department of Energy (DOE) Office of Science by Los Alamos National Laboratory (Contract 89233218CNA000001) and Sandia National Laboratories (Contract DE-NA-0003525). This work was authored in part by the National Renewable Energy Laboratory, operated by Alliance for Sustainable Energy, LLC, for the U.S. Department of Energy (DOE) under Contract No. DE-AC36-08GO28308. The views expressed in the presentation do not necessarily represent the views of the DOE or the U.S. Government. The U.S. Government retains and the publisher, by accepting the article for publication, acknowledges that the U.S. Government retains a nonexclusive, paid-up, irrevocable, worldwide license to publish or reproduce the published form of this work, or allow others to do so, for U.S. Government purposes. R.P. and R.B.C. gratefully acknowledge funding support for film synthesis from the National Science Foundation under award DMR-1809847. R.B.C. also acknowledges funding support for data science and machine learning efforts from the National Science Foundation under award DMR-2045993.  

\section{Competing Interests Statement}
The authors declare no competing interests.

\section{Code Availability Statement}
The code to perform agglomerative clustering and community detection is openly available at \url{https://github.com/pnnl/GraphEM}.

\section{Data Availability Statement}
Data shown is available at \url{https://figshare.com/articles/dataset/Epitaxial_LaFeO_sub_3_sub_SrTiO_sub_3_sub_films_before_and_after_irradiation/27633015}.

\section{Author Contributions}

All authors contributed to the data interpretation and preparation of the article. S.R.S., B.M., J.A.B., and S.A. conceived the study and interpreted the microstructural behavior. A.T-P., M.H., C.D., J.A.B., S.A., and S.R.S. designed and implemented the multi-modal analytics. K.H.Y. conducted SRIM simulations. B.E.M. conducted STEM measurements. L.W., Y.D., R.P., and R.B.C. prepared the thin film samples. E.L. and K.H. performed ion irradiation.

\clearpage

\bibliography{bibliography, spurgeon_refs}

\end{document}